\documentclass[11pt]{article}

\usepackage[a4paper,margin=1in]{geometry}
\usepackage{amsmath,amssymb,amsthm,mathtools,bm}
\usepackage[hidelinks]{hyperref}
\usepackage{enumitem}
\usepackage[title]{appendix}
\usepackage{authblk}
\usepackage{xcolor}

\title{Bond Market Making with a Hit-Ratio Target}
\author{Alexander Barzykin and Axel Ciceri}
\affil{HSBC, 8 Canada Square, Canary Wharf, London E14 5HQ, United Kingdom}
\date{}

\begin{document}
\maketitle

\begin{abstract}
We study OTC bond market making on a size ladder with quadratic inventory penalty and a running target on the dealer's size-weighted hit ratio within a stochastic optimal control approach. We demonstrate that the corresponding reduced Hamilton-Jacobi-Bellman (HJB) equation remains separable by dualizing the hit ratio target term and provides the exact optimal controls through the inverse of the fill-probability function and the Hamiltonian derivative. We then focus on the quadratic approximation à la Bergault \emph{et al.}, which yields a Riccati equation for the inventory curvature while retaining the exact quote map. In its linearized form, this approximation produces explicit quote decompositions into riskless spread, inventory-risk correction, and hit-ratio correction. The formulation is general and applies to multi-bond, multi-client-tier scenarios, with special cases obtained by restricting the targeted tiers, their bond coverage, and their associated targets.
\end{abstract}

\section{Introduction}

Corporate bond trading remains predominantly over-the-counter and dealer-intermediated, even after substantial electronification. In practice, electronification has proceeded less through fully centralized limit-order books than through multi-dealer-to-client request-for-quote (RFQ) protocols, in which a buy-side client solicits a small panel of dealers and typically trades, if at all, with the best respondent. This institutional structure makes bond market making a joint problem of inventory management, counterparty selection, and endogenous execution probability: a dealer must quote competitively enough to win flow, but not so aggressively as to destroy margin or accumulate undesirable inventory. It also means that execution is shaped not only by price, but by the RFQ protocol itself, the number of requested dealers, the response behavior of competitors, and the search technology available to clients.

The RFQ mechanism and competition on dealer platforms is an area of active research. Fermanian \emph{et al.} \cite{FermanianGueantPu2016} provide a foundational empirical model of European corporate-bond RFQs, estimating dealer-quote and client-reservation distributions and explicitly deriving ex-ante hit-ratio functions. Hendershott and Madhavan \cite{HendershottMadhavan2015} analyze the trade-off between search and auction in OTC markets, while Wang \cite{Wang2023} shows that multi-dealer platforms face structural limits in how much price competition they can generate. More recently, Kargar \emph{et al.} \cite{KargarLesterPlanteWeill2025} study sequential search in corporate bonds, and Mar\'{i}n Mart\'{i}nez \emph{et al.} \cite{MarinMartinezArdanzaTrevijanoSabio2026} develop a causal framework for inference on bond multi-dealer-to-client platforms. Taken together, these papers imply that the dealer's hit ratio is not an exogenous statistic, but an equilibrium object shaped by platform design, client search, and rival participation.
In a related foreign exchange market, Oomen \cite{Oomen2017a} provides an economic rationale for limiting the number of liquidity providers exposed to each RFQ and for using routing and ranking rules rather than always inviting the full pool. Barzykin \cite{Barzykin2026} further investigates the role of the win-score promotion gates in aggregator-routed RFQ markets.

Recent studies emphasize the importance of inventory, intermediation capacity, and execution quality in bond dealer markets. Bessembinder \emph{et al.} \cite{BessembinderSpattVenkataraman2020} survey the fixed-income microstructure literature and stress the centrality of OTC frictions, search costs, and slow-moving electronification. In the corporate-bond setting, Bessembinder \emph{et al.} \cite{BessembinderJacobsenMaxwellVenkataraman2018} document the importance of dealer capital commitment, Goldstein and Hotchkiss \cite{GoldsteinHotchkiss2020} study dealer behavior in an illiquid market, and O'Hara \emph{et al.} \cite{OHaraWangZhou2018, OHaraZhou2021} show how execution quality and dealer behavior are altered by electronic RFQ trading. Relationship-based frictions also remain important: Jurkatis \emph{et al.} \cite{JurkatisSchrimpfTodorovVause2023} show that bilateral relationships materially affect trading costs in corporate bonds. For quoting models, this body of evidence argues for combining inventory-sensitive pricing with a realistic description of how RFQs translate into fills.

One mathematical approach to OTC market making employs the stochastic optimal control framework \cite{AvellanedaStoikov2008,GueantLehalleFernandesTapia2013,CarteaJaimungalRicci2014,CarteaJaimungalPenalva2015,Gueant2016}. In particular, Bergault and Gu\'{e}ant \cite{BergaultGueant2021} derive general results for OTC market makers with size-dependent client requests and provide dimensionality-reduction techniques suited to large bond universes \cite{BEGV2021}. Gu\'{e}ant and Manziuk \cite{GueantManziuk2019} address high-dimensional quoting in corporate bonds with a model-based reinforcement-learning approach, while Barzykin \emph{et al.} \cite{BarzykinBergaultGueant2023} extend dealer-market models to include hedging and market impact, thereby capturing the practical transition between pure internalization and external risk transfer. Bergault and Gu\'{e}ant \cite{BergaultGueant2023} further study liquidity dynamics and pricing in RFQ markets using Markov-modulated order-flow models. However, despite the practical salience of hit ratio in bond RFQ workflows, most of this literature treats execution probability in reduced form---typically through quote-dependent intensities or fill probabilities---rather than elevating hit ratio itself to an explicit control target.

This omission matters because hit ratio is not merely a post-trade reporting metric. In industry practice, historical hit ratios are used to select dealers and route RFQs, especially for liquid electronic bond orders, alongside axes, response quality, and relationship information. This suggests a natural modelling question: how should a dealer optimally quote when profitability and inventory control must be balanced not just against a passive fill process, but against an explicit target for win probability or hit ratio? The present paper is aimed at this question. Building on the RFQ and OTC market-making literature, we formulate a stochastic control problem in which quote placement is chosen jointly with a penalty for deviations from a desired hit-ratio objective. This preserves the core inventory-risk logic of OTC market making while incorporating a central practical KPI of electronic bond dealing.

In this work, we follow standard conventions of the optimal market making literature. In particular, the reference process for the instrument is a price, while size is in notional amount. The common practice for corporate bond dealers is to risk manage only the credit exposure of their portfolio (and not the interest rate component), for which the reference process is a credit spread (bp) and sizes are in multiples of the bond's DV01 ($\$ \cdot \text{bp}^{-1})$. Still, our approach yields optimal quotes consistent with the practitioners' objective if one assumes constant DV01 at the timescale of risk management and if the reference price process is interpreted as being driven purely by credit spread changes. These are all reasonable assumptions.

\section{OTC market making and hit-ratio targeting}

\subsection{State dynamics and objective}

We consider a dealer managing a portfolio of bonds offering prices to a group of clients (and/or tiers of clients).
Client tiers are indexed by $\tau\in\mathcal T$ and bonds by ${m\in\mathcal M:=\{1,\dots,M\}}$.
We denote as~$\mathcal A \subseteq \mathcal T$ the subset of client tiers for whom the dealer wishes to target a hit-ratio.
Bond $m$ has mid price process
\begin{equation}
dS^m_t = \sigma^m\, dB_t^m\,,
\end{equation}
where $B_t$ is a standard $M$-dimensional Brownian motion and the covariance matrix is $\Sigma$ (assumed constant on the risk management timescale).
The dealer quotes on both sides of each bond-tier pair across a size ladder indexed by $k=1,\dots,K$, with deterministic sizes $z_k>0$.

Let $s\in\{b,a\}$ denote side, with $b$ for bid and $a$ for ask. For each bond $m$, tier $\tau$, side $s$, and size $k$, RFQ opportunities arrive at exogenous intensity $\lambda_{m,\tau}^{s,k}\geq0$.
\footnote{We require that client tiers have non-trivial total arrival intensity, i.e.~$\displaystyle\sum_{m\in\mathcal M}\displaystyle\sum_{k=1}^K\displaystyle\sum_{s\in\{b,a\}} \lambda_{m,\tau}^{s, k} > 0$, $\forall \tau \in \mathcal T$.}
If the dealer quotes offset $\delta_{m,\tau}^{s,k}\in\mathbb R$ from mid, the fill probability is
\begin{equation}
 f_{m,\tau}^{s,k}(\delta_{m,\tau}^{s,k})\in(0,1)\,,
\end{equation}
so the OTC fill intensity is
\begin{equation}
 \Lambda_{m,\tau}^{s,k}(\delta):=\lambda_{m,\tau}^{s,k}f_{m,\tau}^{s,k}(\delta)\,.
\end{equation}
Let $e_m$ denote the $m$-th canonical basis vector of $\mathbb R^M$. The inventory vector $Q_t\in\mathbb R^M$ and cash process $X_t$ evolve as
\begin{align}
 dQ_t &= \sum_{m\in\mathcal M}\sum_{\tau\in\mathcal T}\sum_{k=1}^K z_k e_m\, \left( d{N_{m,\tau}^{b,k}}_t
      - d{N_{m,\tau}^{a,k}}_t \right),\\
 dX_t &= -\sum_{m,\tau,k}(S^m_t - {\delta^{b,k}_{m,\tau}}_t) z_k\, d{N_{m,\tau}^{b,k}}_t
       +\sum_{m,\tau,k}(S^m_t+ {\delta^{a,k}_{m,\tau}}_t) z_k\,  d{N_{m,\tau}^{a,k}}_t\,,
\end{align}
where $N_{m,\tau}^{s,k}$ are point processes with intensities $\Lambda_{m,\tau}^{s,k}({\delta_{m,\tau}^{s,k}}_t)$.

For any client tier~$\tau\in \mathcal T$ we can define a notional-arrival scale
\begin{equation}
 W_{\tau}:=
 \sum_{m\in\mathcal M} \sum_{s\in\{b,a\}}\sum_{k=1}^K z_k\lambda_{m,\tau}^{s,k},
\end{equation}
and the corresponding size-weighted instantaneous hit ratio
\begin{equation}
 r_\tau(\bm\delta)
 :=
 \frac{1}{W_{\tau}}
\sum_{m\in\mathcal M}  \sum_{s\in\{b,a\}}\sum_{k=1}^K
 z_k\Lambda_{m,\tau}^{s,k}(\delta_{m,\tau}^{s,k}).
\end{equation}
For the targeted client tiers~$\tau\in\mathcal A$, the dealer sets the target levels $r_{\tau}^\star\in(0,1)$.
Given running inventory-risk coefficient $\phi\ge0$, terminal inventory penalty coefficient $\eta\ge0$, and hit-ratio weights $\kappa_{\tau}\ge0$, the dealer solves
\footnote{
In practice, the dealer may wish to target subsets of instruments differently within a given client tier~$\tau$. E.g. bonds in $\mathcal M_1 \subset \mathcal M$ could target hit ratio~$r^\star_\tau = r_1$ while bonds in $\mathcal M_2 \subset \mathcal M$ could target hit ratio $r^\star_\tau = r_2 \ne r_1$.
This can be consistently implemented by splitting the tier $\tau$ into two tiers $\tau_1$ and $\tau_2$.
For $\tau_1$ , we have $\lambda^{s,k}_{m\in\mathcal M_1, \tau_1}>0$ and $\lambda^{s,k}_{m\in\mathcal M_2, \tau_1}=0$, and vice-versa for~$\tau_2$.
Alternatively, the optimization problem~\eqref{eq:main_objective} can be modified to penalize \emph{all} bond/tier-pairs independently, yielding qualitatively similar results.
}
\begin{equation}
\label{eq:main_objective}
 \sup_{\bm\delta}
 \mathbb E\Bigg[
 X_T+Q_T^\top S_T-\frac{\eta}{2}Q_T^\top\Sigma Q_T
 -\int_0^T\left(
 \frac{\phi}{2}Q_t^\top\Sigma Q_t
 +\sum_{\tau\in\mathcal A}\frac{\kappa_{\tau} W_{\tau}}{2}(r_{\tau}(\bm\delta_t)-r_{\tau}^\star)^2
 \right)dt
 \Bigg].
\end{equation}
This specification allows different business targets across targeted tiers with their respective penalty weight $\kappa_{\tau}$.

In practice, dealer performance is typically monitored through an average hit ratio over a reporting window, for example the ratio of the number of won RFQs to the number of received RFQs, possibly after weighting by size or client importance. A fully literal formulation of such a target would therefore require additional state variables describing cumulative request and trade counts, and the objective would involve a ratio of counting processes. While feasible in principle, this would significantly complicate the stochastic-control problem and break much of the tractability sought here. Instead, we penalize deviations of an instantaneous expected hit ratio, constructed from the current request and execution intensities. This should be interpreted as a Markovian surrogate for the desk-level KPI: when RFQs arrive at high frequency and market conditions vary more slowly than individual request outcomes, the business hit ratio over a long window is well approximated by a time average of these local expected hit ratios. In that sense, the present formulation captures the same economic trade-off, balancing margin, inventory, and win probability, while remaining analytically tractable.

\subsection{Reduced HJB, Hamiltonian separation, and exact controls}

Using the standard ansatz, the HJB equation for the reduced value function $u(t,q)$ of time $t$ and inventory vector $q\in\mathbb R^M$ can be written as
\begin{align}
 0={}&\partial_t u(t,q)-\frac{\phi}{2}q^\top\Sigma q \nonumber\\
 &+\sup_{\bm\delta}\Bigg\{
 \sum_{\tau\in\mathcal T}\sum_{m,s,k}
 z_k\Lambda_{m,\tau}^{s,k}(\delta_{m,\tau}^{s,k})\big(\delta_{m,\tau}^{s,k}-\Delta_m^{s,k}u(t,q)\big)
 -\sum_{\tau\in\mathcal A}\frac{\kappa_{\tau} W_{\tau}}{2}(r_{\tau}(\bm\delta)-r_{\tau}^\star)^2
 \Bigg\},
\end{align}
with terminal condition
\begin{equation}
 u(T,q)=-\frac{\eta}{2}q^\top\Sigma q,
\end{equation}
and reduced increments
\begin{equation}
 \Delta_m^{s,k}u(t,q):=\frac{u(t,q)-u(t,q\pm z_ke_m)}{z_k}.
\end{equation}
Hereinafter, the top sign corresponds to bids and the bottom sign to asks, respectively.

Dualizing each targeted penalty term gives
\begin{equation}
 -\frac{\kappa_{\tau}}{2}(r-r_{\tau}^\star)^2=
 \inf_{\xi_{\tau}\in\mathbb R}\left\{\xi_{\tau}(r-r_{\tau}^\star)+\frac{\xi_{\tau}^2}{2\kappa_{\tau}}\right\},
 \qquad \tau\in\mathcal A.
\end{equation}
Applying this identity to the HJB equation then yields
\begin{align}\label{eq:reduced_HJB}
 0={}&\partial_t u(t,q)-\frac{\phi}{2}q^\top\Sigma q + \sum_{\tau\notin \mathcal A}\sum_{m,s,k} z_k  \mathcal H_{m,\tau}^{s,k}\!\left(\Delta_m^{s,k}u \right) \nonumber\\
 &+  \sum_{\tau\in\mathcal A}\inf_{\xi_{\tau}}W_{\tau}
 \left[
 -\xi_{\tau} r_{\tau}^\star
 +\frac{\xi_{\tau}^2}{2\kappa_{\tau}}
 +\frac{1}{W_\tau}\sum_{m, s,k} z_k\,
 \mathcal H_{m,\tau}^{s,k}\!\left(\Delta_m^{s,k}u-\xi_{\tau}\right)
 \right],
\end{align}
where
\begin{equation}\label{eq:hamiltonian_def}
 \mathcal H_{m,\tau}^{s,k}(p):=\sup_{\delta\in\mathbb R}\Lambda_{m,\tau}^{s,k}(\delta)(\delta-p)\,.
\end{equation}

Let~$\tilde\delta_{m,\tau}^{s,k}(p)$ denote the maximizer for the Hamiltonian~\eqref{eq:hamiltonian_def}.
By the envelope theorem, we have
\begin{equation}\label{eq:control_map_general}
 \tilde\delta_{m,\tau}^{s,k}\left(p\right)=
 \left(f_{m,\tau}^{s,k}\right)^{-1}\left(-\frac{\mathcal H_{m,\tau}^{s,k\,\prime}(p)}{\lambda_{m,\tau}^{s,k}}\right)\,.
\end{equation}
We define
\begin{align}
 p_{m,\tau}^{s,k}(t,q):=\left\{
        \begin{array}{ll}
             & \Delta_{m}^{s,k}u(t,q) - \tilde\xi_{\tau}\,, \quad \tau\in \mathcal A \,,\\
&\\
             & \Delta_{m}^{s,k}u(t,q)\,, \quad \tau \notin \mathcal A\,,
        \end{array}
    \right.
\end{align}
where~$\tilde \xi_\tau$ denotes the optimal dual variable associated with~\eqref{eq:reduced_HJB}, i.e.
\begin{equation}\label{eq:tilde_xi_def}
\tilde \xi_\tau \in \arg\min_\xi \bigg\{-\xi r_{\tau}^\star
 +\frac{\xi^2}{2\kappa_{\tau}}
 +\frac{1}{W_\tau}\sum_{m, s,k} z_k\,
 \mathcal H_{m,\tau}^{s,k}\!\left(\Delta_m^{s,k}u-\xi\right)\bigg\}\,,
 \quad \tau\in\mathcal A\,.
 \end{equation}
Then, the optimal policy for bond~$m$,~tier~$\tau$, side~$s$ and size $z_k$, denoted by~$\delta_{m,\tau}^{s,k\,\star}(t,q)$, is
\begin{equation}\label{eq:optimal_controls}
\delta_{m,\tau}^{s,k\,\star}(t,q) = \tilde\delta_{m,\tau}^{s,k}\left(p_{m,\tau}^{s,k}(t,q)\right)\,,
\end{equation}
which is an exact formula via~\eqref{eq:control_map_general}.

The optimal dual variables~$\tilde \xi_\tau$ are obtained for every targeted tier~$\tau\in\mathcal A$ by solving the first-order condition associated with~\eqref{eq:tilde_xi_def}:
\begin{equation}
\label{eq:xi_implicit_general}
 \tilde \xi_{\tau}(t,q)
 =
 \kappa_{\tau}\left(
 r_{\tau}^\star
 +\frac{1}{W_{\tau}}
\sum_{m,k,s} z_k\mathcal H_{m,\tau}^{s,k\,\prime}\!\left(p_{m,\tau}^{s,k}(t,q)\right)
 \right)\,, \quad
 \tau\in\mathcal A.
\end{equation}
This is an implicit characterization because $p_{m,\tau}^{s,k}$ itself depends on~$\tilde \xi_\tau$.

\subsection{Quadratic approximation}\label{subsec:BEGV}

Following Bergault \emph{et al.} (BEGV) \cite{BEGV2021}, we expand the Hamiltonians to the second order in $p$ around 0:
\begin{equation}\label{eq:H_quadratic_expansion}
 \mathcal H_{m,\tau}^{s,k}(p)
 \approx
 \mathcal H_{m,\tau}^{s,k}(0)
 +\mathcal H_{m,\tau}^{s,k\,\prime}(0)\,p
 +\frac12\mathcal H_{m,\tau}^{s,k\,\prime\prime}(0)\,p^2.
\end{equation}
For readability, we work in the side-symmetric setting, and refer to~Appendix~\ref{sec:BEGVNonSym} for the non-symmetric case.
Note that under side symmetry, the bid and ask primitives coincide
\begin{equation}\label{eq:H_symmetry}
\mathcal H_{m,\tau}^{b,k\,(d)}(0)=\mathcal H_{m,\tau}^{a,k\,(d)}(0)\,,
\end{equation}
though we continue to distinguish the sides with index~$s$ for clarity.

We adopt the quadratic ansatz
\begin{equation}
 u(t,q)\approx -\frac12 q^\top A(t)q - C(t),
\end{equation}
where $A(t)$ is symmetric. Then
\begin{equation}
 \Delta_m^{s,k}u(t,q)= e_m^\top A(t) \left( \pm q+\frac12 z_k e_m\right)\,,
\end{equation}
and
\begin{equation}
\label{eq:begv_p_general}
 p_{m,\tau}^{s,k}(t,q)= \left\{
        \begin{array}{ll}
             & e_m^\top A(t) \left( \pm q+\frac12 z_k e_m\right)-\tilde\xi_{\tau}(t,q)\,, \quad \tau\in \mathcal A \,,\\
&\\
             & e_m^\top A(t) \left( \pm q+\frac12 z_k e_m\right)\,, \quad \tau \notin \mathcal A\,.
        \end{array}
    \right.
\end{equation}
At this point, the closure for the optimal dual $\tilde\xi_\tau$, $\tau\in\mathcal A$, can be solved in terms of the value function.
With the quadratic Hamiltonian expansion, we have
\begin{equation}
 \mathcal H_{m,\tau}^{s,k\,\prime}(p)
 \approx
 \mathcal H_{m,\tau}^{s,k\,\prime}(0)+\mathcal H_{m,\tau}^{s,k\,\prime\prime}(0)\,p,
\end{equation}
and thus the $q$-terms cancel bond by bond in~\eqref{eq:xi_implicit_general} after summing bid and ask contributions.
We obtain the inventory-independent result:
\begin{equation}
\label{eq:xi_explicit_general}
 \tilde{\xi}_{\tau}(t) = \tilde \kappa_\tau \, y_\tau(t)\,,
\end{equation}
where
\begin{equation}
\dfrac{1}{\tilde \kappa_\tau} := \dfrac{1}{\kappa_{\tau}} + \dfrac{1}{W_{\tau}}\displaystyle\sum_{m,s,k}^{~} z_k\mathcal H_{m,\tau}^{s,k\,\prime\prime}(0),
\end{equation}
\begin{equation}\label{eq:y_def}
y_\tau(t) :=  r_{\tau}^\star
 +\dfrac{1}{W_{\tau}}\displaystyle\sum_{m,s,k} z_k\mathcal H_{m,\tau}^{s,k\,\prime}(0)
 +\dfrac{1}{2W_{\tau}}\displaystyle\sum_{m,s,k} z_k^2A_{mm}(t)\mathcal H_{m,\tau}^{s,k\,\prime\prime}(0).
\end{equation}

Substituting~\eqref{eq:xi_explicit_general} into the reduced HJB and matching the quadratic terms then yields the matrix Riccati equation
\begin{equation}\label{eq:A_riccati}
 A'(t)=A(t)DA(t)-\phi\Sigma\,,
 \qquad
 A(T)=\eta\Sigma\,,
\end{equation}
with diagonal coefficient matrix
\begin{equation}\label{eq:D_def}
 D:=\mathrm{diag}(d_1,\dots,d_M),
 \qquad
 d_m:=\sum_{\tau,s,k} z_k\mathcal H_{m,\tau}^{s,k\,\prime\prime}(0).
\end{equation}
Thus all available bonds contribute to the inventory curvature, regardless of their belonging to a target tier.

The matrix Riccati equation can be readily solved. Under the natural assumption that $d_m>0$ for every $m$, we have $D \succ0$.
Let $\tilde{t}=T-t$ and
\begin{equation}
G = \left( D^{1/2}\,\phi \Sigma \, D^{1/2} \right)^{1/2}\,.
\end{equation}
Then the solution to~\eqref{eq:A_riccati} is 
\begin{equation}\label{eq:A}
A(t) = D^{-1/2}\, G\,\frac{\sinh \left(\tilde t G\right) + \left( \frac{\eta}{\phi} G \right)\, \cosh \left(\tilde t G\right)}
{\cosh \left(\tilde t G\right) + \left( \frac{\eta}{\phi} G \right)\, \sinh \left(\tilde t G\right)}\, D^{-1/2} \,.
\end{equation}
In the stationary regime, as $\tilde t \to \infty$, the positive semidefinite solution is
\begin{equation}\label{eq:A_infty}
 A =\sqrt{\phi}\,D^{-1/2}(D^{1/2}\Sigma D^{1/2})^{1/2}D^{-1/2}\,.
\end{equation}

A more accurate approach keeps the value-function as derived above in terms of~$A(t)$ in~\eqref{eq:A},
but revisits the dual variable closure~\eqref{eq:xi_implicit_general} with the \emph{exact} nonlinear Hamiltonian response of~\eqref{eq:control_map_general}.
In this way,~\eqref{eq:xi_implicit_general} becomes a scalar fixed-point equation.
Each targeted tier has its own one-dimensional closure (for each fixed inventory state $q$),
which is solved numerically at negligible cost by pre-computing the map $p\mapsto H_{m,\tau}^{k\,\prime}(p)$ from the exact control problem.
While this procedure is no longer analytical, it is still fully in the spirit of BEGV: the value function is kept quadratic, while the nonlinear quote-to-intensity map is preserved exactly.
Alternatively, a local analytical refinement of the explicit static closure can be derived by expanding the exact closure around $q=0$ to second order, as shown in Appendix~\ref{sec:local_quad_xi}.

\subsection{Linearized quotes}\label{subsec:linearized_quotes}

For a closed-form pricing formula, we linearize the control map~\eqref{eq:control_map_general} itself at $p=0$.

Let $\tilde\delta_{m,\tau,0}^{s,k}$ be the size-$z_k$ riskless spread in direction~$s$ for bond $m$ in tier $\tau$, defined by
\begin{equation}
\tilde \delta_{m,\tau,0}^{s,k}:=\arg\max_{\delta}\Lambda_{m,\tau}^{s,k}(\delta)\,\delta\,.
\end{equation}
Then, linearizing the control map~\eqref{eq:control_map_general} yields
\begin{equation}\label{eq:linearized_control_map}
 \tilde\delta_{m,\tau,\star}^{s,k}(p)\approx \tilde\delta_{m,\tau,0}^{s,k}+\frac{p}{c_{m,\tau}^{s,k}},
\end{equation}
where
\begin{equation}
 c_{m,\tau}^{s,k}:=2-\dfrac{f_{m,\tau}^{s,k}(\tilde\delta_{m,\tau,0}^{s,k})f_{m,\tau}^{s,k\,\prime\prime}(\tilde\delta_{m,\tau,0}^{s,k})}{\left(f_{m,\tau}^{s,k\,\prime}(\tilde\delta_{m,\tau,0}^{s,k})\right)^2}\,,
\end{equation}
Moreover, we have
\begin{equation}\label{eq:hamiltonian_coefficients}
 \mathcal H_{m,\tau}^{s,k\,\prime}(0)=-\Lambda_{m,\tau}^{s,k}(\tilde\delta_{m,\tau,0}^{s,k})\,,
 \qquad
 \mathcal H_{m,\tau}^{s,k\,\prime\prime}(0)
 =-\frac{\Lambda_{m,\tau}^{s,k\,\prime}(\tilde\delta_{m,\tau,0}^{s,k})}{c_{m,\tau}^{s,k}}
 =\frac{\Lambda_{m,\tau}^{s,k}(\tilde\delta_{m,\tau,0}^{s,k})}{\tilde\delta_{m,\tau,0}^{s,k} c_{m,\tau}^{s,k}}\,.
\end{equation}
With the linearized control map~\eqref{eq:linearized_control_map}, the linearized optimal quotes follow from~\eqref{eq:optimal_controls} as
\begin{equation}
 \delta_{m,\tau}^{s,k\,\star}(q)
 \approx
  \left\{
        \begin{array}{ll}
             & \tilde\delta_{m,\tau,0}^{s,k}
 +\frac{1}{c_{m,\tau}^{s,k}} e_m^\top A \left( \pm q + \frac12 z_k e_m\right)
 - \frac{1}{c_{m,\tau}^{s,k}}\tilde{\xi}_{\tau}\,, \quad \tau\in \mathcal A \,,\\
&\\
             & \tilde\delta_{m,\tau,0}^{s,k}
 +\frac{1}{c_{m,\tau}^{s,k}} e_m^\top A \left( \pm q + \frac12 z_k e_m\right)\,, \quad \tau \notin \mathcal A\,,
        \end{array}
    \right.
\end{equation}
where~$\tilde{\xi}_{\tau}$ denotes the stationary limit of~$\tilde{\xi}_{\tau}(t)$ in the explicit closure \eqref{eq:xi_explicit_general}, and~$A$ is given in~\eqref{eq:A_infty}.
Hence, all quotes decompose into a riskless-spread component and inventory-risk correction, and quotes for targeted tiers additionally receive a hit-ratio correction.
The sign and magnitude of each $\tilde{\xi}_{\tau}$ summarize whether the corresponding targeted RFQ flow must be subsidized or harvested relative to the riskless-spread benchmark.

\section{Numerical examples}

The general theory covers the main business configurations of interest.
We begin by reviewing the base single bond, single tier scenario, where $M=1$, $\mathcal T=\mathcal A = \{\tau^\star\}$. The inventory is scalar, $A(t)$ is scalar, and the matrix Riccati equation reduces to the classical one-dimensional BEGV equation.

Consider a standard size ladder of $z = (1, 5, 20)$ M notional with baseline RFQ intensities $\lambda_z = (500, 200, 50)$ day$^{-1}$. Win probability is taken to be of a sigmoid shape with $\alpha_z = (2, 1.5, 1)$ and $\beta_z = (2, 1.5, 1)$ bp$^{-1}$.
Target hit ratio is $r^\star = 0.1$, inventory penalty $\phi = 1$ and $\kappa$ is varied.
HJB is solved using explicit Euler scheme\footnote{Numerical solution is implemented in JAX, using pre-calculated Hamiltonian interpolators.} with $T = 1$ days and $q_{\max} = 100$ M, assuming $\eta = 0$.

\begin{figure}[h!]
\centering
\includegraphics[width=0.7\columnwidth]{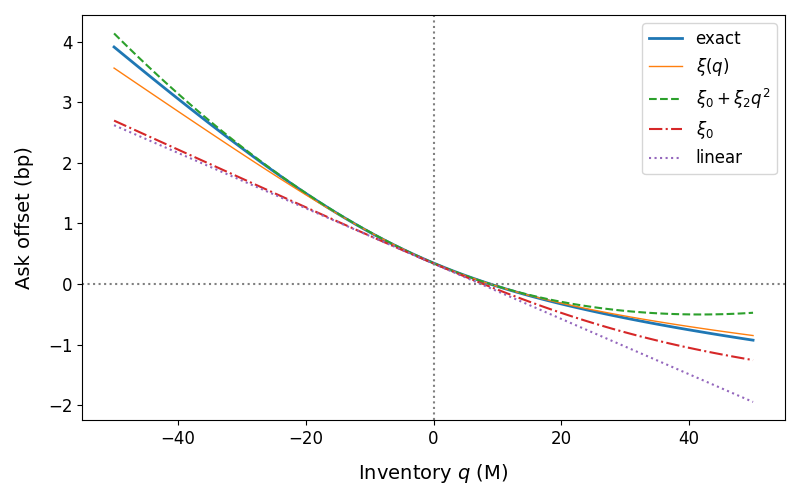}
\caption{
Optimal top of book ($z = 1$) ask offset as a function of inventory $q$
for a single bond, single tier scenario targeting hit ratio of $r^* = 0.1$ with penalty $\kappa = 10$.
Other parameters are given in the text.
Exact numerical solution is compared against different levels of BEGV approximation.
}
\label{fig:ask_of_q_approximations}
\end{figure}

\begin{figure}[h!]
\centering
\includegraphics[width=0.7\columnwidth]{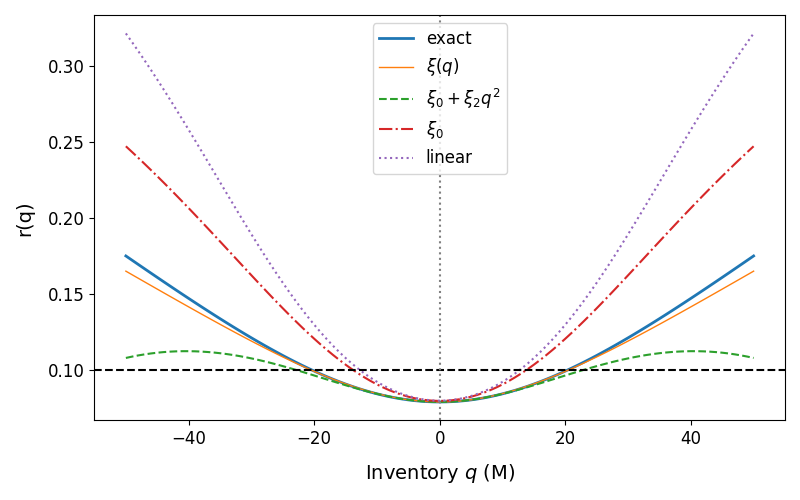}
\caption{
Optimal instantaneous weighted hit ratio as a function of inventory $q$.
Same scenario as in Figure~\ref{fig:ask_of_q_approximations}.}
\label{fig:r_of_q_approximations}
\end{figure}

\begin{figure}[h!]
\centering
\includegraphics[width=0.7\columnwidth]{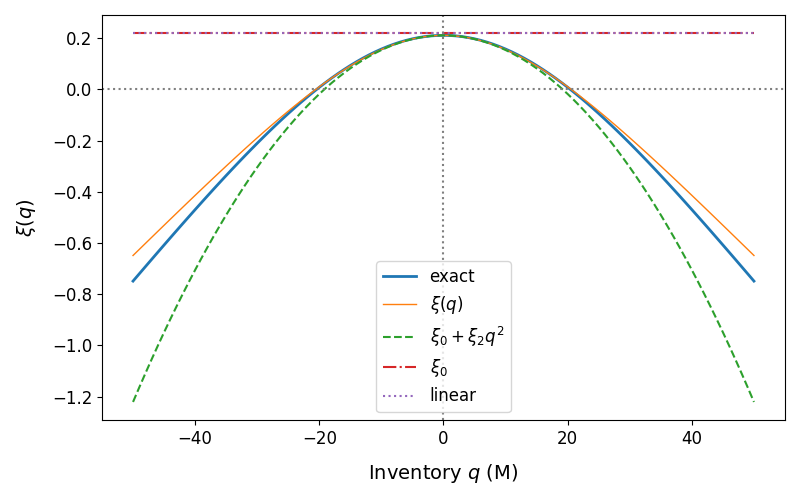}
\caption{
Optimal dual variable as a function of inventory $q$.
Same scenario as in Figure~\ref{fig:ask_of_q_approximations}.}
\label{fig:xi_of_q_approximations}
\end{figure}

First, we compare the exact numerical solution against various levels of BEGV approximation.
Figs.~\ref{fig:ask_of_q_approximations} and \ref{fig:r_of_q_approximations} show state-dependent optimal pricing and instantaneous weighted hit ratio.
For BEGV where the exact nonlinear quote-to-intensity map is preserved and the \emph{exact} closure for~$\tilde\xi$ is used (labeled as $\xi(q)$), the approximation is impressively accurate.
For BEGV where the exact nonlinear quote-to-intensity map is preserved and the \emph{second order analytical} closure is used (labeled as $\xi_0 + \xi_2q^2$), the deviation from the numerical results becomes somewhat significant only for larger inventories.
For BEGV where the exact nonlinear quote-to-intensity map is preserved and the \emph{constant analytical} closure is used (labeled as $\xi_0$), the approximation is good only for small inventories.
Finally, for BEGV where the quote-to-intensity map is linearized, any nonlinearity is lost by construction (though this does not diminish the role of the linear approximation in showcasing meaningful decomposition of pricing components).
Figure~\ref{fig:xi_of_q_approximations} explicitly demonstrates that the exact $\xi(q)$ is indeed strongly nonlinear (approximately quadratic), explaining why the constant closure approximation is poor.

\begin{figure}[h!]
\centering
\includegraphics[width=0.7\columnwidth]{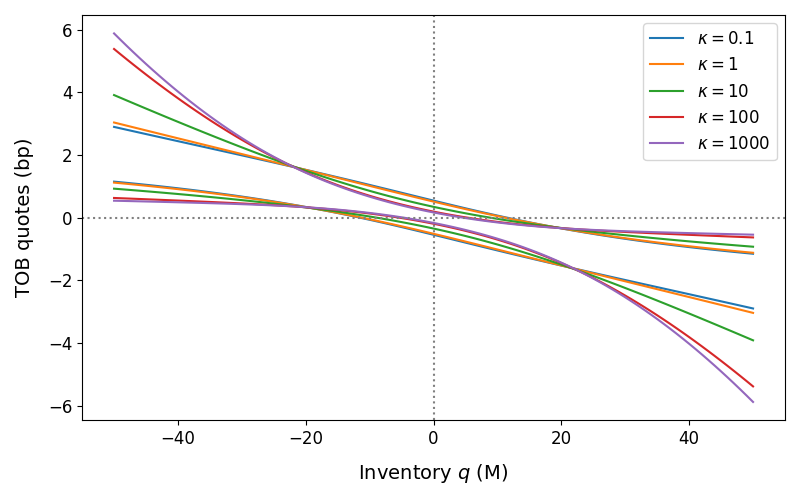}
\caption{
Optimal top of book ($z = 1$) inventory-dependent quotes for a single bond, single tier scenario targeting hit ratio of $r^* = 0.1$ with different value of penalty $\kappa$ (values on chart). Other parameters are given in the text.}
\label{fig:quotes_by_kappa}
\end{figure}

\begin{figure}[h!]
\centering
\includegraphics[width=0.7\columnwidth]{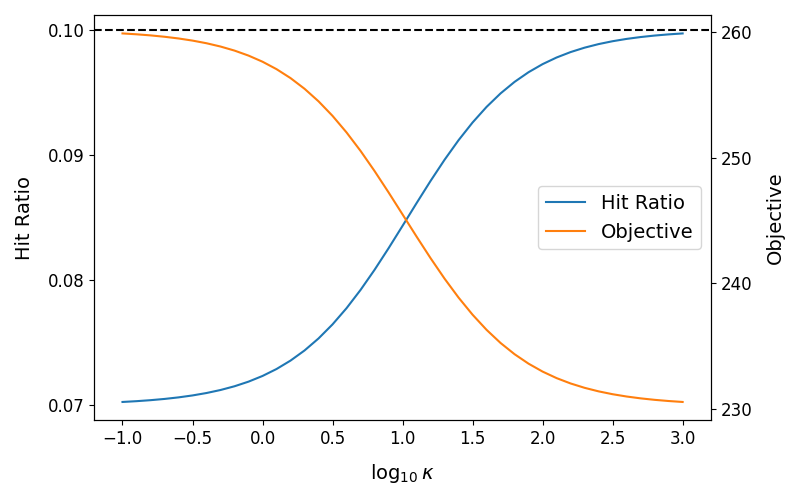}
\caption{
Expected size-weighted hit ratio and overall desk objective \refeq{eq:main_objective} over time horizon $T = 1$ day as functions of hit-ratio penalty coefficient $\kappa$ (log scale).
Dashed line represents the hit-ratio target $r^* = 0.1$.
Other parameters are given in the text.
}
\label{fig:hit_ratio_objective_plot}
\end{figure}

Figure~\ref{fig:quotes_by_kappa} illustrates how optimal pricing responds to increasing hit-ratio penalty.
Since the inventory risk management is also enforced, the pricing is tightened more where the inventory is small.
In order to investigate the effect of targeting on the long-term desk objectives, we propagate the forward inventory law under the optimal stationary policy and evaluate the full mean-field objectives by integrating against the Kolmogorov forward distribution. In particular, for each value of $\kappa$, we first solve the control problem and obtain the optimal stationary quoting policy $q \mapsto \delta^{s, k\,\star}(q)$. This policy induces a continuous-time Markov chain for the inventory process $Q_t$ on the discrete inventory grid $\mathcal Q$. Let $\mu_t = \mathbb P (Q_t = q)$ denote the law of inventory at time $t$ (the probability distribution of $Q_t$ under the chosen policy).
Then $\mu_t$ evolves according to the forward Kolmogorov equation
\begin{equation}
\partial_t \mu_t(q) = \sum_{s, k} \left(
\mathbf 1_{\{q \mp z_k \in \mathcal Q\}} 
\mu_t(q \mp z_k) 
\Lambda^{s,k}\left(\delta^{s,k\,\star}(q \mp z_k) \right)
-\mu_t \Lambda^{s,k}\left(\delta^{s,k\,\star}(q)\right)
\right)
\end{equation}
with initial condition $\mu_0 = \delta_{q_0}$ if the inventory starts from $Q_0 = q_0$.
Once this forward law is known, all performance statistics over a horizon $T$ are computed by averaging the corresponding running quantities against $\mu_t$.
In particular, the expected realized size-weighted hit ratio over the time period, which coincides with the expected time average of the instantaneous size-weighted hit ratio, is given as
\begin{equation}
\mathbb{E} \left[\frac{1}{T} \displaystyle\int_0^T r\left(\bm{\delta^\star}\left(Q_t\right)\right) dt \right]= \frac{
\displaystyle\int_0^T \displaystyle\sum_{q \in \mathcal Q} \mu_t(q) \sum_{s,k} z_k
\Lambda^{s,k}\left(\delta^{s,k\,\star}(q)\right)\,dt
}{
T\displaystyle\sum_{s,k} z_k \lambda^{s,k}
}\,.
\end{equation}
Figure~\ref{fig:hit_ratio_objective_plot} demonstrates how the targeting penalty $\kappa$ controls the expected hit ratio.
The dealer faces a multi-objective optimization aiming to maximize P\&L while minimizing inventory risk and deviation from hit-ratio target, so spending more on target generally implies less P\&L.

It is much more realistic from business perspective that the dealer would target only particular clients while keeping risk management of the rest of the franchise intact. For instance, we consider one bond, a single targeted tier~$\tau^\star$ and multiple non-targeted tiers, i.e.~$M=1$, $\mathcal T=\{\tau_0\,, \tau_1\,, \cdots\,,  \tau^\star\}$ and $\mathcal A=\{\tau^\star\}$.
The scalar Riccati coefficient $D$ sums the contributions of all tiers because they all affect inventory risk, whereas the dual closure \eqref{eq:xi_explicit_general} only involves the targeted tier. Thus untargeted tiers contribute to the inventory curvature but not directly to the hit-ratio correction. This scenario is still one-dimensional and can be readily solved numerically.

\begin{figure}[h!]
\centering
\includegraphics[width=0.7\columnwidth]{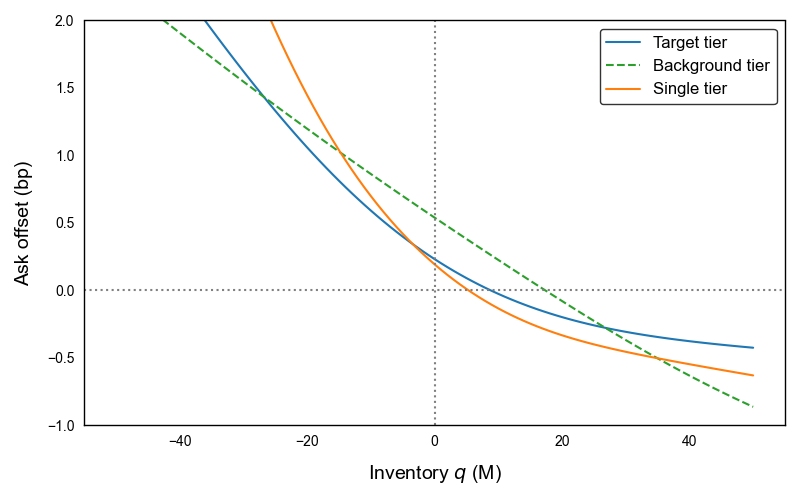}
\caption{
Optimal top of book ($z = 1$) inventory-dependent quotes for a single bond, single tier scenario  (solid orange) and a single bond, two tier scenario for target tier (solid blue) and background tier (dashed).
Target hit ratio $r^* = 0.1$ in both cases, and $\kappa = 100$.
Other parameters are given in the text.
}
\label{fig:pricing_2tier}
\end{figure}

\begin{figure}[h!]
\centering
\includegraphics[width=0.7\columnwidth]{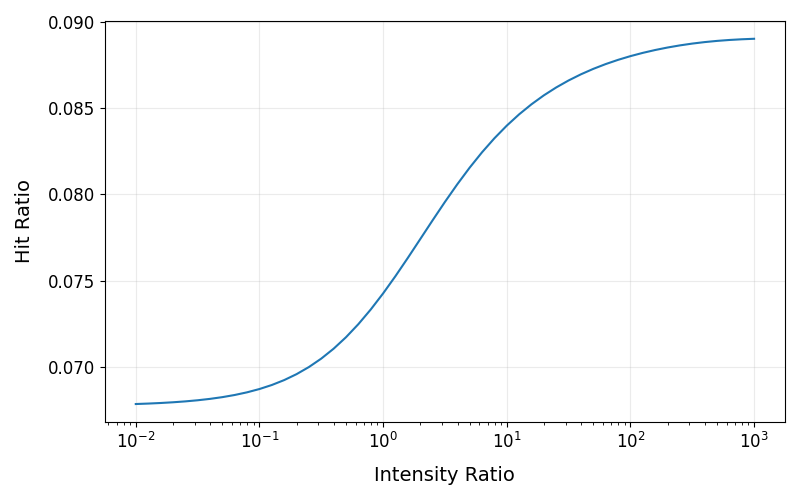}
\caption{
Expected size-weighted hit ratio over time horizon $T = 1$ day as a function of intensity ratio of background to targeted tier.
Target hit ratio $r^* = 0.1$, $\kappa = 100$.
Other parameters are given in the text.
}
\label{fig:hit_ratio_2tier}
\end{figure}

Figure~\ref{fig:pricing_2tier} demonstrates this case for~$\mathcal T=\{\tau_0\,,  \tau^\star\}$.
It shows the influence of the ``background tier''~$\tau_0$ with exactly the same parameters on pricing.
The background tier takes part in inventory risk management and thus reduces pressure on the targeted tier.
As a result, pricing becomes less skewed and risk-neutral spread (at $q=0$) widens.
Expected hit ratio also improves for a given $\kappa$ when the background intensity share is increased, as shown in Figure~\ref{fig:hit_ratio_2tier}.
In this experiment, we set $\lambda_z = (50, 20, 5)$ day$^{-1}$ for the targeted tier, i.e. 10 times smaller in order to cover a wider realistic range of intensity ratios.

\begin{figure}[h!]
\centering
\includegraphics[width=0.7\columnwidth]{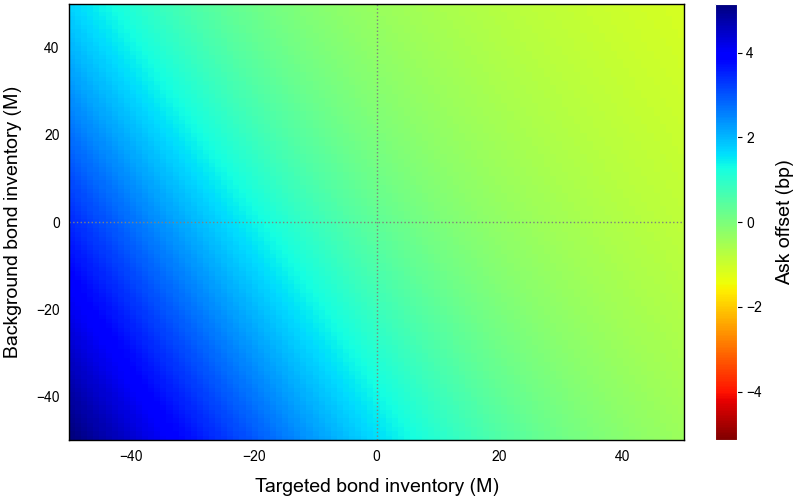}
\caption{
Optimal top of book ($z = 1$) quotes for the bond in the targeted tier, in the presence of a correlated bond in the non-targeted tier.
Target hit ratio $r^* = 0.1$, $\kappa = 10$, $\rho = 0.8$.
Other parameters are given in the text.
}
\label{fig:pricing_2bond}
\end{figure}

\begin{figure}[h!]
\centering
\includegraphics[width=0.7\columnwidth]{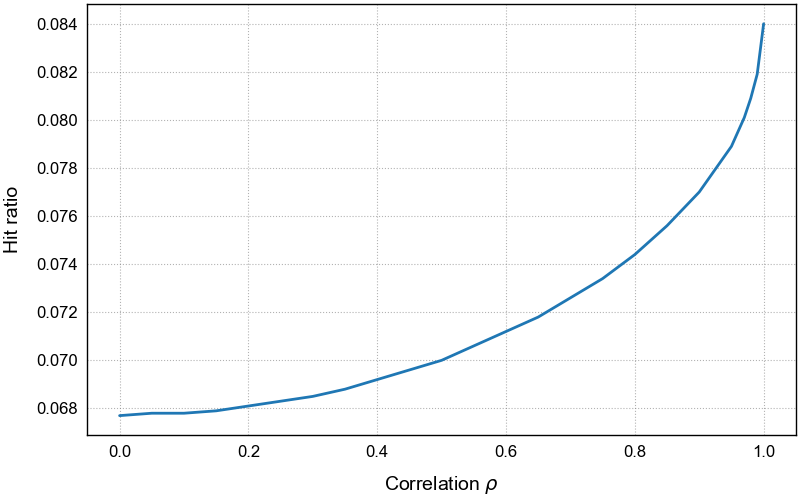}
\caption{
Expected size-weighted hit ratio of the bond in the targeted tier over time horizon $T = 1$ day as a function of bond correlation coefficient.
Target hit ratio $r^* = 0.1$, $\kappa = 10$.
Other parameters are given in the text.
}
\label{fig:hit_ratio_2bond}
\end{figure}

Another practical scenario considers a bond portfolio $m\in\mathcal M$ with one targeted tier and one non-targeted tier, i.e. $\mathcal T=\{\tau_0, \tau^\star\}$ with $\mathcal A=\{\tau^\star\}$.
For simplicity, we can assume that the targeted tier sees flow only in bond~$m^\star$, i.e.~$\lambda_{m, \tau^\star}^k=0\,,\forall m\neq m^\star$.
 Then $A(t)$ is matrix-valued, and the quote on bond $m$ depends on the full inventory vector through $e_m^\top A(t)q$. The dual closure for the targeted tier only involves the bond~$m^\star$, but its coefficient $(A(t))_{m^\star m^\star}$ is influenced by the entire covariance matrix $\Sigma$ and by all quoting opportunities. Figure~\ref{fig:pricing_2bond} illustrates the pricing surface for the targeted bond~$m^\star$ in the presence of another correlated bond in the background tier with the same parameters. Since this latter background bond contributes to risk management, the skew of the bond in the targeted tier is significantly reduced when there is an offsetting inventory in the background bond. Figure~\ref{fig:hit_ratio_2bond} further demonstrates that the expected hit ratio is improved by the presence of the correlated background bond. In this case, we set $\lambda_z = (50, 20, 5)$ day$^{-1}$ for the bond in the targeted tier, i.e.~10 times smaller than for the bond in the background tier, in order to amplify the effect.

\section{Concluding remarks}

We have studied a market-making problem in which the dealer not only manages inventory risk and trading profitability, but also explicitly targets hit ratio. Motivated by RFQ markets, where win probability is a central business KPI, we introduced a quadratic penalty for deviations from prescribed hit-ratio targets and embedded it into a stochastic-control framework. This led to an HJB equation in which the usual quoting trade-off is enriched by an additional control channel reflecting the value of improving fill probability. A key structural result is that, after dualization of the hit-ratio penalty, the problem retains a tractable Hamiltonian form. In particular, the HJB can be written with decoupled Hamiltonians, which preserves much of the analytical convenience of classical market-making models while accommodating targeted execution-quality constraints. This makes it possible to treat single-bond, multi-tier, and multi-bond settings within a unified formulation, and more generally to assign distinct hit-ratio targets to different RFQ segments. On the numerical side, we solved the resulting control problems with explicit schemes and showed that hit-ratio targeting can materially alter optimal pricing and inventory dependent capture rates. At the same time, we derived a BEGV-style quadratic approximation, together with refined closures for the dual variable associated with hit-ratio targeting. These approximations proved to be remarkably accurate in the numerical experiments, suggesting that they provide a useful reduced-form description of the full control problem. Although the discussion was motivated in part by bond RFQ workflows, the analysis is not bond-specific. The same modelling ideas apply more broadly to RFQ markets in which dealers quote to heterogeneous client flow and are judged not only on spread capture and inventory management, but also on execution probability. In that sense, the present framework offers a tractable way to integrate a practically important business objective into the mathematical theory of OTC market making.

\section*{Acknowledgment}
The authors are grateful to Richard Anthony, Joshua Freeland and Austin Cottrell (HSBC) for support throughout the project and valuable discussions.
The views expressed are those of the authors and do not necessarily reflect the views or practices at HSBC.

\appendix

\begin{appendices}

\section{Quadratic approximation for non-symmetric trade intensities}
\label{sec:BEGVNonSym}

In Section~\ref{subsec:BEGV}, we applied the BEGV framework to derive an approximation for the value function and optimal dual variables under the assumption of symmetric trade intensities.
In this section, we highlight the key differences for the arbitrary, non-symmetric case. With asymmetric intensities, the bid/ask Hamiltonian functions~$\mathcal H_{m,\tau}^{b,k}(p)$,~$\mathcal H_{m,\tau}^{a,k}(p)$, do not coincide.
Thus, the equality of the bid/ask primitives~\eqref{eq:H_symmetry} in the quadratic expansion~\eqref{eq:H_quadratic_expansion} no longer holds.
Relatedly, the ansatz for the value function must be extended to admit a linear term in~$q$:
\begin{equation}
 u(t,q)\approx -\frac12 q^\top A(t)q  - B(t)^\top q - C(t),
\end{equation}
where $A(t)$ is symmetric. Then
\begin{equation}
 \Delta_m^{s,k}u(t,q)= e_m^\top A(t) \left( \pm q+\frac12 z_k e_m\right) \pm B_m(t).
\end{equation}
Evaluating the first-order condition \eqref{eq:xi_implicit_general} for the optimal dual variable now yields
\begin{equation}
\label{eq:xi_explicit_general_asym}
 \tilde{\xi}_{\tau}(t,q) =  \tilde \kappa_\tau \, y_\tau(t)
 +
 \dfrac{\tilde \kappa_\tau}{W_{\tau}} \sum_m h_{m,\tau}^{12}\left[e^\top_{m}A(t)q + B_m(t)\right] \,, \quad \tau \in \mathcal{A}\,,
\end{equation}
where
\begin{align}
 h_{m,\tau}^{ij} ={} & \sum_k (z_k)^i \left(\mathcal H_{m,\tau}^{b, k\,(j)}(0) - \mathcal H_{m,\tau}^{a, k\,(j)}(0)\right)\,.
\end{align}
The first term is the symmetric closure~\eqref{eq:xi_explicit_general} in the main text,
while the second term arises from asymmetric trading opportunities $h_{m,\tau}^{12} \neq 0$.
Note that the $q$-dependence is generically present.

Matching the quadratic terms in the reduced HJB yields the matrix Riccati equation
\begin{equation}
 A'(t)=A(t){\mathcal D}A(t)-\phi\Sigma,
\label{eq:matrix_riccati}
 \qquad
 A(T)=\eta\Sigma\,,
\end{equation}
where
\begin{equation}\label{eq:K_def}
 \mathcal D := 
D - \displaystyle \sum_{\tau\in\mathcal A} \frac{\tilde \kappa_\tau}{W_\tau} 
	\sum_{m,m\prime}h_{m,\tau}^{12}\,h_{m\prime,\tau}^{12} e_m e_{m\prime}^\top
\,.
\end{equation}

We first show that $\mathcal{D}$ is positive definite under the standard assumptions.
Let
\[
 d_{m,\tau}:=\sum_{s,k} z_k\mathcal H_{m,\tau}^{s,k\,\prime\prime}(0)\,, \quad D_\tau :=\mathrm{diag}(d_{1,\tau},\dots,d_{M,\tau})\,,
\]
and
\[
\mathcal D_\tau= \left\{
        \begin{array}{ll}
             & D_\tau - \frac{\tilde{\kappa}_\tau}{W_\tau} \sum_{m,m'} h_{m,\tau}^{12} h_{m',\tau}^{12} e_m e_{m'}^\intercal\,, \quad \tau\in \mathcal A \,,\\
&\\
             & D_\tau \,, \quad \tau \notin \mathcal A\,,
        \end{array}
    \right.
\]
so that
\[
D=\sum_{\tau\in\mathcal T}D_\tau\,, \qquad \mathcal D=\sum_{\tau\in\mathcal T}\mathcal D_\tau\,.
\]
For $\tau\notin\mathcal A$ one already has $\mathcal D_\tau=D_\tau\succ0$, so it remains only to show that $\mathcal D_\tau\succ0$ for $\tau\in\mathcal A$.
Since each Hamiltonian is strictly convex, $\mathcal H^{s,k\,\prime\prime}_{m,\tau}(0) > 0$, we have $|h_{m,\tau}^{12}| < d_{m,\tau}$\,.
It follows that, for every non-zero vector $x \in \mathbb{R}^M\setminus \{0\}$,
\begin{equation}\label{eq:inequality_h_d}
\left(\sum_m h_{m,\tau}^{12}x_m\right)^2 < \left( \sum_m d_{m,\tau}\right) x^\intercal\, D_\tau\, x\,.
\end{equation}
Moreover, for~$\tau \in \mathcal A$, we have
\[
\frac{\tilde{\kappa}_\tau}{W_\tau} = \frac{1}{W_\tau / \kappa_\tau + \sum_m d_{m,\tau}}\,,
\]
which is strictly positive for non-trivial~$\kappa_\tau>0$.
Therefore, using~\eqref{eq:inequality_h_d},
\[
x^\intercal \left( D_\tau - \frac{\tilde{\kappa}_\tau}{W_\tau} \sum_{m,m'} h_{m,\tau}^{12} h_{m',\tau}^{12} e_m e_{m'}^\intercal\right) x > \frac{W_\tau/\kappa_\tau\,\,x^\intercal\, D_\tau\,x}{W_\tau/\kappa_\tau + \sum_m d_{m,\tau}} > 0\,, 
\]
and so~$\mathcal D_\tau \succ0$ for~$\tau \in \mathcal A$\,.
We conclude that $\mathcal{D}  \succ0$.

Having established the positive-definiteness of~$\mathcal{D}$, the solution to \eqref{eq:matrix_riccati} is
\begin{equation}
A(t) = \mathcal D^{-1/2}\, \mathcal{G}\,
\frac{\sinh \left((T-t) \mathcal{G}\right) + \left(\frac{\eta}{\phi} \mathcal G\right) \cosh\left( (T-t) \mathcal{G}\right)}
{\cosh \left( (T-t) \mathcal{G}\right) + \left(\frac{\eta}{\phi} \mathcal G\right) \sinh\left((T-t) \mathcal{G}\right)}
\, \mathcal D^{-1/2} \,,
\end{equation}
where
\begin{equation}
\mathcal G := \left( \mathcal D^{1/2}\,\phi \Sigma \, \mathcal D^{1/2} \right)^{1/2}\,,
\end{equation}
and so in the stationary regime, as $T \to \infty$, we have
\begin{equation}\label{eq:A_asym_stationary}
 A=\sqrt{\phi}\,\mathcal D^{-1/2}(\mathcal D^{1/2}\Sigma \mathcal D^{1/2})^{1/2}\mathcal D^{-1/2}.
\end{equation}

With~$A(t)$ given, the matching of the linear terms in the reduced HJB yields a non-homogeneous linear ODE for~$B(t)$:
\begin{equation}
- B'(t) + A(t)\mathcal D B(t)= Y(t),
 \qquad
 B(T)=0,
\end{equation}
where~$Y(t)$ is the vector given by
\begin{equation}\label{eq:Y_def}
  Y(t) := A(t)\sum_{m}\Bigg[
  - \sum_{\tau\in\mathcal T} \left( h_{m,\tau}^{11} + \frac 12  A_{mm}(t) h_{m,\tau}^{22} \right)
  + \sum_{\tau\in\mathcal A} \tilde \kappa_\tau y_\tau(t) h_{m,\tau}^{12} 
  \Bigg]e_m\,,
\end{equation}
and~$y_{\tau}(t)$ is defined in \eqref{eq:y_def}.
The solution is given in integral form as
\begin{align}
B(t) ={} \mathcal D^{-1/2} &\left[\cosh \left( (T-t)\mathcal G\right)  + \left(\frac{\eta}{\phi}\mathcal G\right) \sinh \left( (T-t)\mathcal G\right) \right]^{-1} \nonumber \\
  \times \int_t^T & \left[\cosh \left( (T-s)\mathcal G\right)  + \left(\frac{\eta}{\phi}\mathcal G\right) \sinh \left( (T-s)\mathcal G\right) \right] \mathcal D^{1/2} Y(s)~ds\,.
\end{align}
In the stationary limit, $T \to \infty$, the integral resolves as
\begin{equation}\label{eq:B_asym_stationary}
B = \mathcal D^{-1/2} \int_0^\infty e^{-s \mathcal G} \mathcal D^{1/2}Y~ ds = \mathcal D^{-1/2} \mathcal G^{-1} \mathcal D^{1/2} Y\,,
\end{equation}
where~$Y$ is the constant vector obtained by taking the stationary limit of~$Y(t)$ in~\eqref{eq:Y_def}.

The corresponding linearized optimal quotes in the stationary limit are
\begin{equation}
\delta_{m,\tau}^{s,k\,\star}(q)
 \approx
  \left\{
        \begin{array}{ll}
             &\tilde \delta_{m,\tau,0}^{s,k}
 +\frac{1}{c_{m,\tau}^{s,k}} e_m^\top A \left( \pm q + \frac{z_k}{2} e_m\right) \pm \frac{1}{c_{m,\tau}^{s,k}} B_m
 - \frac{1}{c_{m,\tau}^{s,k}}\tilde{\xi}_{\tau}(q)\,, \quad \tau\in \mathcal A \,,\\
&\\
             &\tilde \delta_{m,\tau,0}^{s,k}
 +\frac{1}{c_{m,\tau}^{s,k}} e_m^\top A \left( \pm q + \frac{z_k}{2} e_m\right)  \pm \frac{1}{c_{m,\tau}^{s,k}}B_m \,, \quad \tau \notin \mathcal A\,,
        \end{array}
    \right.
\end{equation}
where~$\tilde{\xi}_{\tau}(q)$ denotes the stationary limit of~$\tilde{\xi}_{\tau}(t,q)$ in~\eqref{eq:xi_explicit_general_asym}.

\section{Local quadratic expansion of dual variable}
\label{sec:local_quad_xi}

A local analytical refinement of the explicit constant closure \eqref{eq:xi_explicit_general}
can be obtained by expanding each exact scalar closure \eqref{eq:xi_implicit_general} around $q=0$. 
We work in the side-symmetric case, and for convenience we drop the tilde when referring to the optimal dual $\tilde\xi_\tau$.

For a fixed $t$ and $\tau \in \mathcal A$ define
\begin{equation}
F_{\tau}(\xi,q;t) :=  - r_{\tau}^\star + \frac{\xi}{\kappa_{\tau}} - \frac{1}{W_{\tau}} \sum_{m,s,k} z_k  \mathcal H_{m,\tau}^{s,k\, \prime} \big( p_{m,\tau}^{s,k}(t,q,\xi) \big),
\end{equation}
where, under the quadratic BEGV ansatz for the targeted tiers,
\begin{equation}
p_{m,\tau}^{s,k}(t,q,\xi) =  e_m^\top A(t)\left(  \pm  q + \frac{z_k}{2}  e_m \right)- \xi\,.
\end{equation}
The refined closure~\eqref{eq:xi_implicit_general} is the implicit equation
\begin{equation}
F_{\tau}(\xi_{\tau}(t,q),q;t) = 0.
\end{equation}

Let $\xi_{\tau,0}(t):=\xi_{\tau}(t,0)$ and define
\begin{equation}
 \zeta_{m,\tau}^k(t):=\frac{z_k}{2} A_{mm}(t)-\xi_{\tau,0}(t)
 \qquad \text{for } \tau\in\mathcal A.
\end{equation}
Then $\xi_{\tau,0}(t)$ solves $F_{\tau}(\xi_{\tau,0}(t), 0; t) = 0$. If $\partial_\xi F_{\tau}(\xi_{\tau,0}(t), 0; t) \ne 0$, the implicit function theorem guarantees that there exists a locally smooth function $q \mapsto \xi_{\tau}(t,q)$ solving the closure near $q=0$. The required derivatives are:
\begin{equation}
\partial_\xi F_{\tau}(\xi,q;t) = \frac{1}{\kappa_{\tau}} + \frac{1}{W_{\tau}} \sum_{m,s,k} z_k \mathcal H_{m,\tau}^{s,k\,\prime\prime}(p_{m,\tau}^{s,k}),
\end{equation}
and hence at $q=0$,
\begin{equation}
\partial_\xi F_{\tau}(\xi_{\tau,0}(t),0;t) = \frac{1}{\kappa_{\tau}} + \frac{1}{W_{\tau}} \sum_{m, s, k} z_k \mathcal H_{m,\tau}^{s, k\,\prime\prime}(\zeta_{m,\tau}^k).
\end{equation}

Next, taking partial derivatives with respect to $q$ while holding $\xi$ fixed,
\begin{equation}
\nabla_q F_{\tau}(\xi,q;t) = -\frac{1}{W_{\tau}} \sum_{m,k} z_k \left( \mathcal H_{m,\tau}^{b,k\,\prime\prime}(p_{m,\tau}^{b,k}) - \mathcal H_{m,\tau}^{a,k\,\prime\prime}(p_{m,\tau}^{a,k})\right) A(t) e_m.
\end{equation}
At $q=0$, the bid and ask arguments coincide, so $\nabla_q F_{\tau}(\xi_{\tau,0}(t),0;t), 0; t) = 0$. Therefore $\nabla_q \xi_{\tau}(t, 0) = 0$, which is why the first nontrivial correction is quadratic (in side-symmetric case). Differentiating once more with respect to $q$, one gets
\begin{equation}
D_q^2 F_{\tau}(\xi_{\tau,0}(t),0;t) = -\frac{1}{W_{\tau}} \sum_{m,s,k} z_k \mathcal H_{m,\tau}^{s,k\,\prime\prime\prime}(\zeta_{m,\tau}^k(t)) A(t) e_m e_m^\top A(t).
\end{equation}
Since $\nabla_q \xi_{\tau}(t, 0)=0$, the second-order implicit differentiation formula simplifies to
\begin{equation}
D_q^2 \xi_{\tau}(t,0) = - \frac{D_q^2 F_{\tau}(\xi_{\tau,0}(t),0;t)}{\partial_\xi F_{\tau}(\xi_{\tau,0}(t),0;t)}.
\end{equation}
Thus writing
\begin{equation}
\label{eq:xi_quadratic_general}
 \xi_{\tau}(t,q)
 \approx
 \xi_{\tau}(t)+\frac12 q^\top  B_\tau(t) q,
\end{equation}
where~${\xi}_{\tau}(t)$ is the zeroth-order approximation~\eqref{eq:xi_explicit_general},
the Hessian coefficient is
\begin{equation}
\label{eq:B_xi_general}
B_\tau(t)
 =
 \frac{1}{W_{\tau}}
 \frac{
 \displaystyle\sum_{m,s,k} z_k\mathcal H_{m,\tau}^{s,k\,\prime\prime\prime}(\zeta_{m,\tau}^k(t))\,A(t)e_me_m^\top A(t)
 }{
 \dfrac{1}{\kappa_{\tau}}
 +\dfrac{1}{W_{\tau}}\displaystyle\sum_{m,s,k} z_k\mathcal H_{m,\tau}^{s,k\,\prime\prime}(\zeta_{m,\tau}^k(t))
 },
 \qquad \tau\in\mathcal A.
\end{equation}

\end{appendices}


\begin{thebibliography}{99}

\bibitem{AvellanedaStoikov2008}
M.~Avellaneda and S.~Stoikov,
\newblock High-frequency trading in a limit order book,
\newblock \emph{Quant.~Finance}, 2008, {\bfseries 8}, 217--224.

\bibitem{BarzykinBergaultGueant2023}
A.~Barzykin, P.~Bergault, and O.~Gu\'{e}ant,
\newblock Algorithmic market making in dealer markets with hedging and market impact,
\newblock \emph{Math.~Finance}, 33(1):41--79, 2023.

\bibitem{Barzykin2026}
A.~Barzykin,
\newblock Win-score promotion gates in aggregator-routed RFQ markets: A two-tier stochastic control model,
\newblock \emph{arXiv:2603.10569}, 2026.

\bibitem{BergaultGueant2021}
P.~Bergault and O.~Gu\'{e}ant,
\newblock Size matters for OTC market makers: General results and dimensionality reduction techniques,
\newblock \emph{Math.~Finance}, 31(1):279--322, 2021.

\bibitem{BEGV2021}
P.~Bergault, D.~Evangelista, O.~Guéant and D.~Vieira,
\newblock Closed-form approximations in multi-asset market making.
\newblock \emph{Appl.~Math.~Finance}, 2021, {\bfseries 28}, 101--142.

\bibitem{BergaultGueant2023}
P.~Bergault and O.~Gu\'{e}ant,
\newblock Liquidity dynamics in RFQ markets and impact on pricing,
\newblock \emph{arXiv:2309.04216}, 2023.

\bibitem{BessembinderJacobsenMaxwellVenkataraman2018}
H.~Bessembinder, S.~Jacobsen, W.~Maxwell, and K.~Venkataraman,
\newblock Capital commitment and illiquidity in corporate bonds,
\newblock \emph{J.~Finance}, 73(4):1615--1661, 2018.

\bibitem{BessembinderSpattVenkataraman2020}
H.~Bessembinder, C.~Spatt, and K.~Venkataraman,
\newblock A survey of the microstructure of fixed-income markets,
\newblock \emph{J.~Financ.~Quant.~Anal.}, 55(1):1--45, 2020.

\bibitem{CarteaJaimungalRicci2014}
\'{A}.~Cartea, S.~Jaimungal, and J.~Ricci,
\newblock Buy low, sell high: A high frequency trading perspective,
\newblock \emph{SIAM~J.~Financial~Math.}, 5(1):415--444, 2014.

\bibitem{CarteaJaimungalPenalva2015}
\'{A}.~Cartea, S.~Jaimungal, and J.~Penalva,
\newblock \emph{Algorithmic and High-Frequency Trading},
\newblock Cambridge University Press, 2015.

\bibitem{FermanianGueantPu2016}
J.-D.~Fermanian, O.~Gu\'{e}ant, and J.~Pu,
\newblock The behavior of dealers and clients on the European corporate bond market: The case of Multi-Dealer-to-Client platforms,
\newblock \emph{Mark.~Microstruct.~Liq.}, 2(03n04):1750004, 2016.

\bibitem{GoldsteinHotchkiss2020}
M.~A.~Goldstein and E.~S.~Hotchkiss,
\newblock Providing liquidity in an illiquid market: Dealer behavior in U.S. corporate bonds,
\newblock \emph{J.~Financ.~Econ.}, 135(1):16--40, 2020.

\bibitem{GueantLehalleFernandesTapia2013}
O.~Guéant, C.-A.~Lehalle and J.~Fernandez-Tapia,
\newblock Dealing with the inventory risk: a solution to the market making problem.
\newblock \emph{Math.~Financ.~Econ.}, 2013, {\bfseries 7}, 477--507.

\bibitem{Gueant2016}
O.~Guéant,
\newblock \emph{The Financial Mathematics of Market Liquidity: From optimal execution to market making},
\newblock Chapman and Hall/CRC: Boca Raton, FL, 2016.

\bibitem{GueantManziuk2019}
O.~Gu\'{e}ant and I.~Manziuk,
\newblock Deep reinforcement learning for market making in corporate bonds: Beating the curse of dimensionality,
\newblock \emph{Appl.~Math.~Finance}, 26(5):387--452, 2019.

\bibitem{HendershottMadhavan2015}
T.~Hendershott and A.~Madhavan,
\newblock Click or call? Auction versus search in the over-the-counter market,
\newblock \emph{J.~Finance}, 70(1):419--447, 2015.

\bibitem{JurkatisSchrimpfTodorovVause2023}
S.~Jurkatis, A.~Schrimpf, K.~Todorov, and N.~Vause,
\newblock Relationship discounts in corporate bond trading,
\newblock \emph{BIS Working Papers}, No.~1140, 2023.

\bibitem{KargarLesterPlanteWeill2025}
M.~Kargar, B.~Lester, S.~Plante, and P.-O.~Weill,
\newblock Sequential search for corporate bonds,
\newblock \emph{Federal Reserve Bank of Philadelphia Working Paper} 25-08, 2025.

\bibitem{MarinMartinezArdanzaTrevijanoSabio2026}
P.~Mar\'{i}n Mart\'{i}nez, S.~Ardanza-Trevijano, and J.~Sabio,
\newblock Causal interventions in bond multi-dealer-to-client platforms,
\newblock \emph{PLOS ONE}, 21(1):e0341369, 2026.

\bibitem{OHaraWangZhou2018}
M.~O'Hara, Y.~Wang, and X.~(Alex) Zhou,
\newblock The execution quality of corporate bonds,
\newblock \emph{J.~Financ.~Econ.}, 130(2):308--326, 2018.

\bibitem{OHaraZhou2021}
M.~O'Hara and X.~(Alex) Zhou,
\newblock The electronic evolution of corporate bond dealers,
\newblock \emph{J.~Financ.~Econ.}, 140(2):368--390, 2021.

\bibitem{Oomen2017a}
R.~Oomen,
\newblock Execution in an aggregator.
\newblock \emph{Quant.~Finance}, 2017, {\bfseries 17}, 383--404.

\bibitem{Wang2023}
C.~Wang,
\newblock The limits of multi-dealer platforms,
\newblock \emph{J.~Financ.~Econ.}, 149(3):434--450, 2023.

\end{thebibliography}
\end{document}